\documentclass[10pt]{article}

\usepackage[top=1.0in,right=1.0in,left=1.0in,bottom=1.75in]{geometry}
\usepackage{amssymb,amsbsy}
\usepackage{amsmath}
\usepackage{amsthm}
\usepackage{graphics,graphicx}
\usepackage[T1]{fontenc}
\usepackage{color}
\usepackage{subfigure}
\usepackage[affil-it,auth-sc]{authblk}
\usepackage{stmaryrd}

\usepackage[numbers,sort&compress]{natbib}

\newcommand{\beq}{\begin{equation}}
\newcommand{\eeq}{\end{equation}}

\newcommand{\beqar}{\begin{eqnarray}}
\newcommand{\eeqar}{\end{eqnarray}}
\newcommand{\bit}{\begin{itemize}}
\newcommand{\eit}{\end{itemize}}
\newcommand{\benum}{\begin{enumerate}}
\newcommand{\eenum}{\end{enumerate}}
\newcommand{\barr}{\begin{array}}
\newcommand{\earr}{\end{array}}

\newcommand\eq[1]{(\ref{#1})}

\def\XXint#1#2#3{{\setbox0=\hbox{$#1{#2#3}{\int}$}
   \vcenter{\hbox{$#2#3$}}\kern-.5\wd0}}

\def\b0{\mbox{\boldmath $0$}}

\def\f0{\ensuremath{\mathbb{O}}}

\newcommand{\mG}{\ensuremath{\mathcal{G}}}

\newcommand{\sign}{\mathop{\mathrm{sign}}}

\title{Dispersion and localization in structured Rayleigh beams}

\author[1]{A. Piccolroaz\footnote{Corresponding author: e-mail: roaz@ing.unitn.it; phone: +39\,0461\,282583.}}
\author[2]{A.B. Movchan}

\affil[1]{Dipartimento di Ingegneria Meccanica e Strutturale, Universit\`a di Trento, Italy}
\affil[2]{Department of Mathematical Sciences, University of Liverpool, U.K.}

\date{}

\begin{document}

\maketitle

\begin{abstract}
\noindent

This paper brings a comparative analysis between dynamic models of couple-stress elastic materials and 
structured Rayleigh beams on a Winkler foundation. Although physical phenomena have different  physical origins, the underlying equations appear to be similar, and hence mathematical models have a lot in common.  In the present work, our main focus is on the analysis of dispersive waves, band-gaps and localized waveforms in structured Rayleigh beams. 
The Rayleigh beam theory includes the effects of rotational inertia which are neglected in the Euler--Bernoulli beam theory. This makes the approach applicable to higher frequency regimes. Special attention is given to waves in pre-stressed Rayleigh beams on elastic foundations.

\end{abstract}



%

\section{Introduction and analogy between waves in Rayleigh beams and couple-stress elastic materials}
\label{sec01}

Bloch waves in structured media have received a significant attention in models of photonic and phononic crystals, that encompass problems of electro-magnetism, optics, acoustics and more recently elasticity \cite{Yablonovitch1987,John1987,Kushwaha1993,Sigalas1993}.
An important feature of elastic waves, even in an isotropic case, is the presence of two types of waves linked to dilatation and shear, respectively.

More recently, there was a significant interest generated by studies of micropolar media and couple-stress materials
(see, for example, \cite{Morini2013b,Engelbrecht2013,Zisis2014,Gourgiotis2014}).
In particular, the Mindlin's approach \cite{Mindlin1964} leads to additional higher-order derivatives in the governing equations. Engelbrecht et al.\ \cite{Engelbrecht2005} follow Mindlin's interpretation of a micro-structure as a polycrystal, whose micro-elements are taken as deformable cells. In the limit when the cells are rigid, their approach would lead to the Cosserat model. It was noted that the higher-order model leads to the novel dispersion properties of waves supported by such a micro-structured medium. 
We also note a formal equivalence between constrained Cosserat model and the couple-stress model of structured media (see, for example, \cite{Koiter1964,Mindlin1962}), and a further analogy with the theory of Rayleigh beams which accounts for rotational inertia, which will be discussed in the text below.   

When rotational inertia is neglected in analysis of flexural waves, this is fully appropriate in the long wave approximations and is well adopted to the Euler-Bernoulli beam theory. Dispersion and filtering of elastic waves in structured prestressed Euler--Bernoulli beams were considered in \cite{Gei2009}, 
which includes analysis of band-gap shift, defect-induced annihilation and localized modes.
It is also known that the couple-stress effects are neglected in the classical models of linear elasticity. 

It appears that there is a mathematical underlying framework, which applies both to the couple-stress approach in elasticity as well as flexural waves in beam models accounting for the effects of rotational inertia.

The Rayleigh beam theory is used here to account for the rotary motion of beam elements. This approach also allows for the description of flexural waves  at high frequency ranges \cite{Graff1991}. 
In the Rayleigh beam theory, the assumptions regarding the geometry of the deformation and the material properties remain unchanged, so that the rotation 
$\phi$ of the cross sections is not an independent parameter, but it is {\em constrained} to the transverse displacement $v$ by the relation $\phi = v'$. 
However, in writing the equations of motion, both the translational inertia and the rotational inertia of beam elements are taken into account, so that
\begin{equation}
V' = \beta v + \rho A \ddot{v}, \quad M' = V - P v' - \rho I \ddot{v}',
\end{equation}
where $V$ is the internal shear force, $M$ the internal bending moment, $\beta$ the stiffness of a Winkler type elastic foundation, $P$ the prestress, 
$\rho$ the mass density, $A$ the area of the cross-section, and $I$ the area moment of inertia of the cross-section. 

Combining the equations of motion with the {\em constitutive} equation $M = -EI v''$, where $EI$ is the bending stiffness, we obtain the governing equation 
for the transverse motion $v$ of a homogeneous prestressed Rayleigh beam resting on an elastic foundation of the Winkler type
\begin{equation}
\label{ray}
EI\, v'''' - P\, v'' + \rho A\, \ddot{v} - \rho I\, \ddot{v}'' + \beta v = 0.
\end{equation}

We note that this physical problem has a formal mathematical analogy with the problem of shear wave propagation in couple-stress elastic materials.
In fact, the dynamic equation governing the anti-plane motion in a couple-stress elastic material is given by \cite{Mishuris2012}
\begin{equation}
\label{cstress}
G \ell^2\, \Delta^2 w - 2G\, \Delta w + 2\rho\, \ddot{w} - \frac{J}{2}\, \Delta \ddot{w} = 0,
\end{equation}
where $w$ is the out-of-plane displacement; the shear modulus $G$ and the mass density $\rho$ are the classical {\em macroscopic} material parameters, whilst the characteristic length 
$\ell$ and the micro-rotational inertia $J$ are the generalized parameters connected with the {\em microstructure}. 
Clearly, equation (\ref{cstress}) is the two-dimensional analogue of the Rayleigh beam equation (\ref{ray}) with $\beta = 0$. 

Both equations (\ref{ray}) and (\ref{cstress}) are not  classical wave equations, because of the fourth-order terms. In the case of the Rayleigh beam, the coefficients near fourth-order terms include the bending stiffness $EI$ and the rotational inertia $\rho I$, showing that they are essentially linked to the geometrical properties of the structural element, in particular the area moment of inertia of the cross-section.
On the other hand, in the case of a couple-stress material, the fourth-order terms are proportional to the microstructural parameters $G\ell^2$ and $J$. In both cases, these higher-order terms are responsible for the dispersive character of wave propagation.

In several studies on wave propagation in generalized continua \cite{Mindlin1964,Gourgiotis2013,Morini2014}, it has been shown that elastic waves can be made non-dispersive for special values of the rotational inertia. In particular, for shear waves in couple stress materials described by eq.\ (\ref{cstress}), this value is given by $J = 2 \rho \ell^2$ \cite{Mishuris2012}. By analogy, we obtain that flexural waves in a Rayleigh beam are non-dispersive for a special value of the prestress, namely $P = EA$.

\begin{figure}[!htcb]
\centering
\includegraphics[width=15cm]{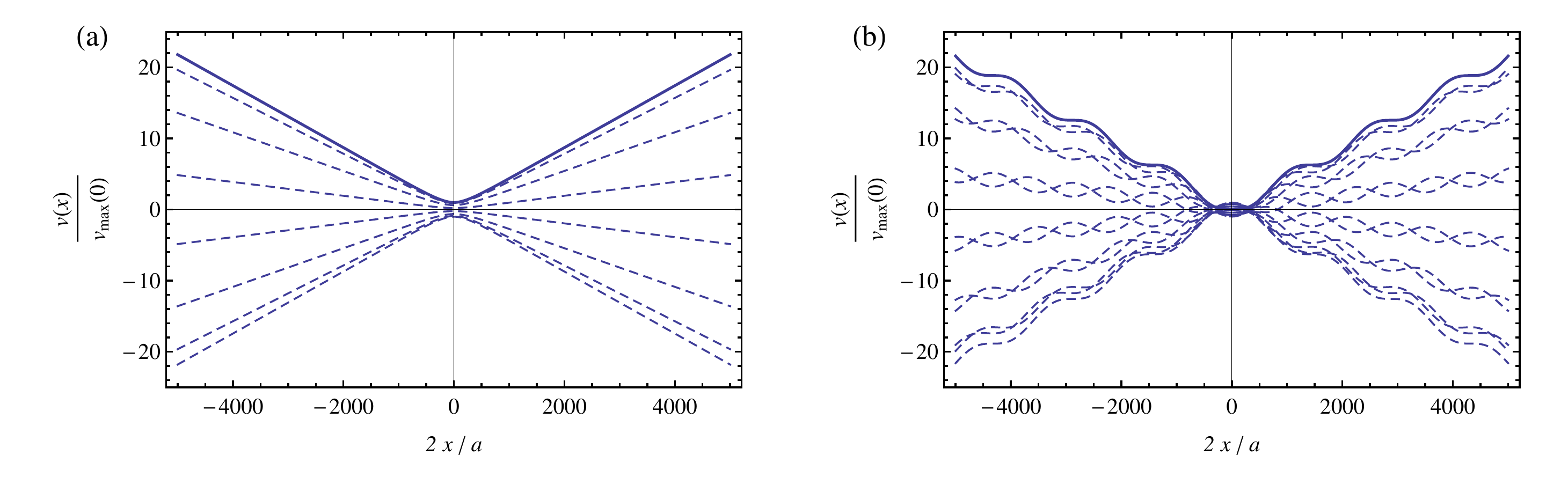}
\caption{\footnotesize Flexural vibrations produced by a time-harmonic point force applied at the origin to a Rayleigh beam on the Winkler foundation in the special resonant cases when the solutions have a linear growth at infinity (see Section \ref{sec04}, formulae (\ref{GF40}) and (\ref{GF41})). We plot several profiles captured at different times within the period of the oscillatory force. A beam with circular cross section is assumed having the following properties: Young modulus 210 GPa, mass density 7.85$\cdot$10$^3$ kg/m$^3$, foundation stiffness 2.64 MPa, radius of the cross section 0.01 m. The angular frequency is 1034.4 s$^{-1}$. Transversal displacements $v(x)$ are normalised with respect to the maximum value at the origin $v_\text{max}(0)$, whereas the longitudinal position $x$ is normalised with respect to the radius of the cross-section $a$. (a) Tensile prestress equal to 1.32 kN. (b) Compressive prestress equal to -1.19 kN. }
\label{fig00}
\end{figure}

The key features characterising a dynamic response of a Rayleigh beam on the Winkler foundation are linked to the presence of exponential and quasi-polynomial terms in the representation of flexural displacements. In particular,  exponential terms account for the high gradient regions, and of course they can be seen in the representation of dynamic Green's functions.
For the purpose of illustration, we show in Fig.~\ref{fig00} the graphs of flexural displacements $v(x)$ produced by the time-harmonic point force applied to the Rayleigh beam at the origin. 
These represent the special resonant cases when the solution has a linear growth at infinity; the oscillation in Fig.~\ref{fig00}b is produced as a result of a compressive prestress. 

The structure of the paper is as follows. Section \ref{sec02} presents a detailed account of dispersion properties of flexural waves in a homogeneous prestressed Rayleigh beam supported by a Winkler foundation.
In Section \ref{sec03} we consider a periodic multi-phase medium, with Bloch waves representing quasi-periodic solutions of the equation of motion. Analysis of band gaps and filtering properties is included in this section. Section \ref{sec04} includes analytical closed form representations for dynamic Green's functions and their derivatives (concentrated couples), that describe the full range of admissible localised and propagating wave forms. Finally, in Section \ref{sec05}, we construct a special class of so-called quasi-periodic Green's functions, required to study a dynamic response of periodic systems of masses placed along the Rayleigh beam; this also includes periodic system of bodies with given rotational inertia. The coupling between the rotational action and transverse motion of masses appears to be important, which is discussed in that section. 
The analysis is generic and our systematic study is applicable to flexural systems such as plates and shells, especially for the cases where rotational inertia appears to provide a significant contribution.

\section{Dispersion properties of a homogeneous prestressed Rayleigh beam on an elastic foundation}
\label{sec02}

The dispersion relation for a homogeneous prestressed Rayleigh beam, governed by \eq{ray}, is obtained by assuming that a sinusoidal signal $v = \xi \exp i(kx - \omega t)$ is propagating in the beam, which gives
\begin{equation}
\label{disp}
\omega = \sqrt{\frac{k^2r^2(k^2r^2 + \overline{P}) + \overline{B}}{R(k^2r^2 + 1)}},
\end{equation}
where $r$ is the radius of inertia of the beam cross-section
\begin{equation}
r = \sqrt{\frac{I}{A}},
\end{equation}
$\overline{P}$ and $\overline{B}$ are the dimensionless prestress and foundation stiffness, respectively,
\begin{equation}
\overline{P} = \frac{Pr^2}{EI} = \frac{P}{EA}, \quad
\overline{B} = \frac{\beta r^4}{EI} = \frac{\beta I}{EA^2},
\end{equation}
and $R$ is the normalized inertia term, having the dimension of a squared time
\begin{equation}
R = \frac{\rho r^2}{E} = \frac{\rho I}{EA}.
\end{equation}
Correspondingly, the phase velocity $c$ and the group velocity $c_g$ are given by
\begin{equation}
c = \frac{r}{\sqrt{R}} \sqrt{\frac{k^2r^2 (k^2r^2 + \overline{P}) + \overline{B}}{k^2r^2(k^2r^2 + 1)}}, \quad
c_g = \frac{r^2}{R} \frac{k^4r^4 + 2k^2r^2 + \overline{P} - \overline{B}}{c(k^2r^2 + 1)^2}.
\end{equation}

\subsection{Asymptotics limits and the action of prestress}

The solution of the dispersion equation (\ref{disp}) has the following properties: 
\begin{enumerate}
\item In the long-wavelength limit, $kr \to 0$, the angular frequency behaves as
\begin{equation}
\omega = \sqrt{\frac{\overline{B}}{R}} + \frac{\overline{P} -\overline{B}}{2\sqrt{R\overline{B}}} k^2r^2 + O(k^4r^4).
\end{equation}
\item In the short-wavelength limit, $kr \to \infty$, the angular frequency behaves as
\begin{equation}
\omega = \frac{1}{\sqrt{R}} kr + \frac{\overline{P} - 1}{2\sqrt{R}(kr)} + O(k^{-3}r^{-3}).
\end{equation}
\item For a compressive prestress, $\overline{P}$ < 0, (or tensile prestress, satisfying $\overline{P} < \overline{B}$) the angular frequency displays a minimum $R\omega_{\text{min}}^2 = 2\sqrt{1 + \overline{B} - \overline{P}} + \overline{P} - 2$ for $k^2r^2 = \sqrt{1 + \overline{B} - \overline{P}} - 1$. Consequently, buckling occurs when both phase and group velocities are zero, $c = 0$ and $c_g = 0$. These conditions are met for a compressive prestress $\overline{P}$ satisfying $\overline{P}_{\text{buckl}} = -2\sqrt{\overline{B}}$ with $k_{\text{buckl}}^2r^2 = \sqrt{\overline{B}}$.\footnote{
The buckling load can be obtained also from the static equation $EI v'''' - P v'' + \beta v = 0$, which admits a non-trivial and bounded solution only for compressive prestress, $\overline{P} < 0$, and non-negative determinant of the characteristic equation, $\Delta = \overline{P}^2 - 4\overline{B} \geq 0$. 
}
\end{enumerate}

\subsection{Dispersion diagrams for Rayleigh beam versus Euler-Bernoulli beam}

\begin{figure}[!htcb]
\centering
\includegraphics[width=16cm]{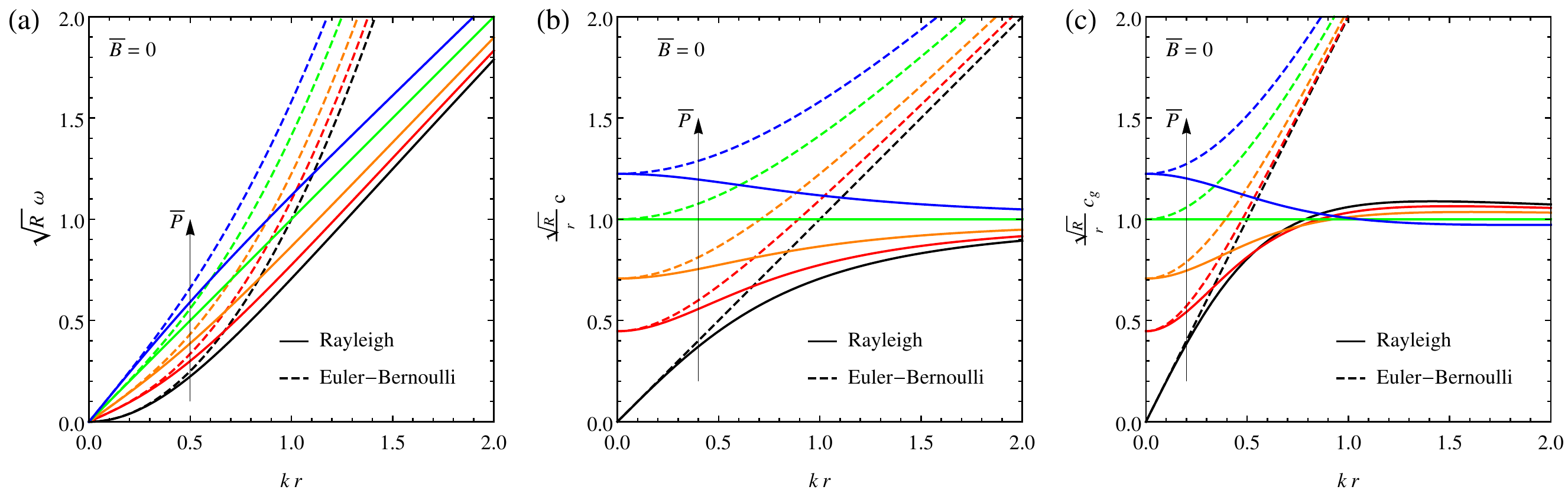}
\caption{\footnotesize Dispersion diagrams for a homogeneous prestressed Rayleigh beam. (a) Normalised angular frequency $\sqrt{R}\omega$ vs.\ normalised wave number $k r$; (b) Normalised phase velocity $\sqrt{R} c / r$ vs.\ normalised wave number $k r$; (c) Normalised group velocity $\sqrt{R} c_g / r$ vs.\ normalised wave number $k r$. Curves are plotted for five values of prestress $\overline{P} = \{0,0.2,0.5,1,1.5\}$. Dispersion diagrams for a homogeneous prestressed Euler--Bernoulli beam are also plotted for comparison (dashed lines). Note that for a Rayleigh beam and for prestress $\overline{P} = 1$ waves are non-dispersive.}
\label{fig01}
\end{figure}

Three diagrams in Fig.~\ref{fig01} include the graphs of normalised frequency, phase and group velocities of flexural waves as functions of $kr$ for a homogeneous prestressed Rayleigh beam and $\overline{B} = 0$.
Curves are plotted for five values of normalised prestress $\overline{P} = \{0,0.2,0.5,1,1.5\}$. Dispersion diagrams for a homogeneous prestressed Euler--Bernoulli beam are also plotted for comparison (dashed lines).
In the absence of the elastic foundation, the Rayleigh and the Euler-Bernoulli beams deliver a similar dynamic response in the low frequency regimes. However, as the frequencies rise, the behaviour of elastic waves propagating along the Rayleigh and the Euler-Bernoulli beams is notably different. In contrast with the Euler--Bernoulli beam, where dispersive flexural waves show quadratic growth of frequency as a function of $k r$, the waves in the Rayleigh beam become asymptotically non-dispersive in the high-frequency regimes, as their group and phase velocities tend to a constant as $k r  \to \infty$. The action of prestress is especially important in the low frequency regime, as the higher positive prestress leads to an increase in the phase and group velocities of flexural waves. Contribution from prestress appears to be small at higher frequencies. 

\begin{figure}[!htcb]
\centering
\includegraphics[width=16cm]{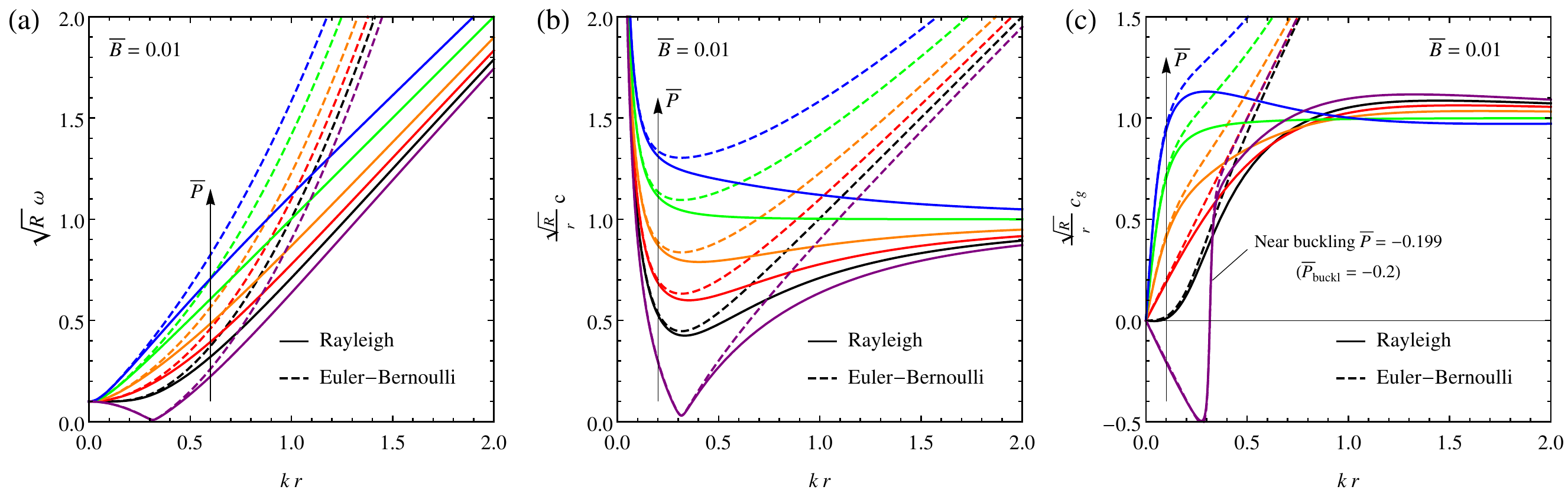}
\caption{\footnotesize Dispersion diagrams for a homogeneous prestressed Rayleigh beam on an elastic foundation ($\overline{B} = 0.01$). (a) Normalised angular frequency $\sqrt{R}\omega$ vs.\ normalised wave number $k r$; (b) Normalised phase velocity $\sqrt{R} c / r$ vs.\ normalised wave number $k r$; (c) Normalised group velocity $\sqrt{R} c_g / r$ vs.\ normalised wave number $k r$. Curves are plotted for six values of prestress $\overline{P} = \{-0.199,0,0.2,0.5,1,1.5\}$. Dispersion diagrams for a homogeneous prestressed Euler--Bernoulli beam are also plotted for comparison (dashed lines). Note that for compressive prestress $\overline{P} = -0.199$ the beam is near the onset of buckling, $\overline{P}_{\text{buckl}} = -0.2$.}
\label{fig02}
\end{figure}

Similarly, in Fig.~\ref{fig02}, we show the graphs of normalised frequency, phase and group velocities for elastic configurations that include the Winkler foundation ($\overline{B} = 0.01$). Curves are plotted for six values of normalised prestress $\overline{P} = \{-0.199,0,0.2,0.5,1,1.5\}$. Although it does not produce significant change at high frequencies, the changes observed in the low frequency regimes are dramatic. It is important to note non-monotonic behaviour of $\sqrt{R}\omega$ as a function of $kr$ both for the Rayleigh and the Euler--Bernoulli beams; the local minimum point, where the group velocity is zero corresponds to a standing wave supported by the action of compressive prestress. In particular, when the magnitude of the compressive prestress reaches the level, such that the point of minimum corresponds to the zero frequency, we observe buckling of the beam - in this very special case both Rayleigh and Euler--Bernoulli beams respond in the same manner.
Note that the lower dispersion curve in Fig.~\ref{fig02}a corresponds to compressive prestress $\overline{P} = -0.199$ and the beam is near the onset of buckling. 
For the chosen value of foundation stiffness $\overline{B} = 0.01$, the buckling prestress is $\overline{P}_{\text{buckl}} = -0.2$. As $\overline{P} \to \overline{P}_{\text{buckl}}$, both phase and group velocities tend to zero for the same value of the wave number, $kr = 0.01^{1/4}$, corresponding to a buckled configuration with wavelength $\lambda/r = 2\pi/(kr) = 19.8692$. 

\section{Dispersion and band-gaps for a structured Rayleigh beam}
\label{sec03}

In this section we analyse the dispersion properties of Bloch-Floquet flexural waves within a prestressed structured periodic Rayleigh beam on an elastic foundation of the Winkler type.
Analytical estimates are derived  for the boundaries of stop bands, i.e. the intervals of  frequencies at which the waves are evanescent. Exponentially localised waveforms occur at these frequencies when a point force is applied to the Rayleigh beam, which highlights the importance of our results.

\begin{figure}[!htcb]
\centering
\includegraphics[width=8cm]{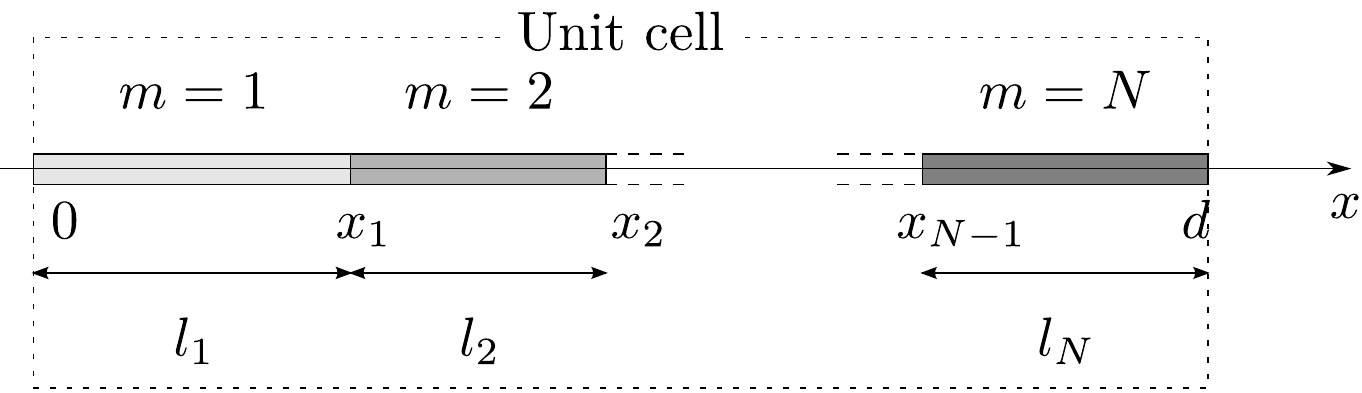}
\caption{\footnotesize Unit cell.}
\label{fig03a}
\end{figure}

For each of the phases $m$ ($m = 1,...,N$)  within the unit cell of the periodic structure, Fig.~\ref{fig03a}, the time-harmonic transverse displacement $v_m$ satisfies the following ordinary differential equation
\begin{equation}
\label{gov}
E_m I_m v_m'''' - (P - \rho_m I_m \omega^2) v_m'' + (\beta - \rho_m A_m \omega^2) v_m = 0.
\end{equation}
The internal bending moment $M_m$ and the internal shear force $V_m$ are given by
\begin{equation}
M_m = -E_m I_m v_m'', \quad V_m = -E_m I_m v_m''' + (P - \rho_m I_m \omega^2) v_m',
\end{equation}
respectively.

The flexural displacements are sought in the form
\begin{equation}
\label{form}
v_m = C^{(m)} \exp(i k^{(m)} x),
\end{equation}
The substitution of (\ref{form}) into (\ref{gov}) yields the following equations for the circular frequency $\omega$
\begin{equation}
\label{quartic}
(k^{(m)} r_m)^4 + (\overline{P}_m - R_m\omega^2) (k^{(m)} r_m)^2 + \overline{B}_m - 
R_m\omega^2 = 0,
\end{equation}
where we have introduced the following dimensionless parameters
\begin{equation}
\overline{P}_m = \frac{Pr_m^2}{E_mI_m}, \quad
\overline{B}_m = \frac{\beta r_m^4}{E_mI_m},
\end{equation}
and the normalized quantity
\begin{equation}
R_m = \frac{\rho_m A_m r_m^4}{E_mI_m}.
\end{equation}
The quartic (\ref{quartic}) admits the following four roots
\begin{equation}
\label{roots}
k^{(m)}_{1,2,3,4} = \pm \frac{1}{r_m} \sqrt{-\frac{\overline{P}_m - R_m\omega^2}{2} 
\pm \sqrt{\frac{(\overline{P}_m - R_m\omega^2)^2}{4} + (R_m\omega^2 - \overline{B}_m)}}.
\end{equation}
Consequently, the general solution for the transverse displacement $v_m$ within each phase is given by
\begin{equation}
v_m(x) = \sum_{p=1}^{4} C_p^{(m)} \exp(ik_p^{(m)}x), 
\end{equation}
The $4N$ constants $C_p^{(m)}$ can be obtained by imposing the junction conditions at each internal interface of the 
elementary block ($m = 1, ..., N-1$), namely continuity of displacement, rotation, bending moment, and the shear force,
\begin{equation}
\label{eq1}
v_m(x_m) = v_{m+1}(x_m), 
\end{equation}
\begin{equation}
v_m'(x_m) = v_{m+1}'(x_m),
\end{equation}
\begin{equation}
-E_mI_m v_m''(x_m) = -E_{m+1}I_{m+1} v_{m+1}''(x_m), 
\end{equation}
\begin{equation}
\label{eq2}
-E_m I_m v_m'''(x_m) + (P - \rho_m I_m \omega^2) v_m'(x_m) = 
-E_{m+1} I_{m+1} v_{m+1}'''(x_m) + (P - \rho_{m+1} I_{m+1} \omega^2) v_{m+1}'(x_m).
\end{equation}
The remaining four equations follow from enforcing the Bloch-Floquet conditions at the boundary of the elementary block,
\begin{equation}
\label{eq3}
v_1(0) = v_N(d) \exp(-iKd),
\end{equation}
\begin{equation}
v_1'(0) = v_N'(d) \exp(-iKd),
\end{equation}
\begin{equation}
-E_1 I_1 v_1''(0) = -E_N I_N v_N''(d) \exp(-iKd),
\end{equation}
\begin{equation}
\label{eq4}
-E_1 I_1 v_1'''(0) + (P - \rho_1 I_1 \omega^2)v_1'(0) = 
[-E_N I_N v_N'''(d) + (P - \rho_N I_N \omega^2)v_N'(d)] \exp(-iKd),
\end{equation}
where $K$ is the Bloch parameter. Equations (\ref{eq1})--(\ref{eq2}) and (\ref{eq3})--(\ref{eq4}) 
provide a homogeneous linear system for the $4N$ unknown constants $C_p^{(m)}$, $p = 
1,...,4$, $m = 1,...,N$, so that the vanishing of the determinant of the associated matrix yields the 
dispersion equation for the structured Rayleigh beam.

\paragraph{Analytical estimates of band gap positions.}

A simple formula can be obtained for the estimation of band gap positions by considering homogeneous beams with physical characteristics equal to each of the constituents of the periodic beam.

\begin{figure}[!htcb]
\centering
\includegraphics[width=12cm]{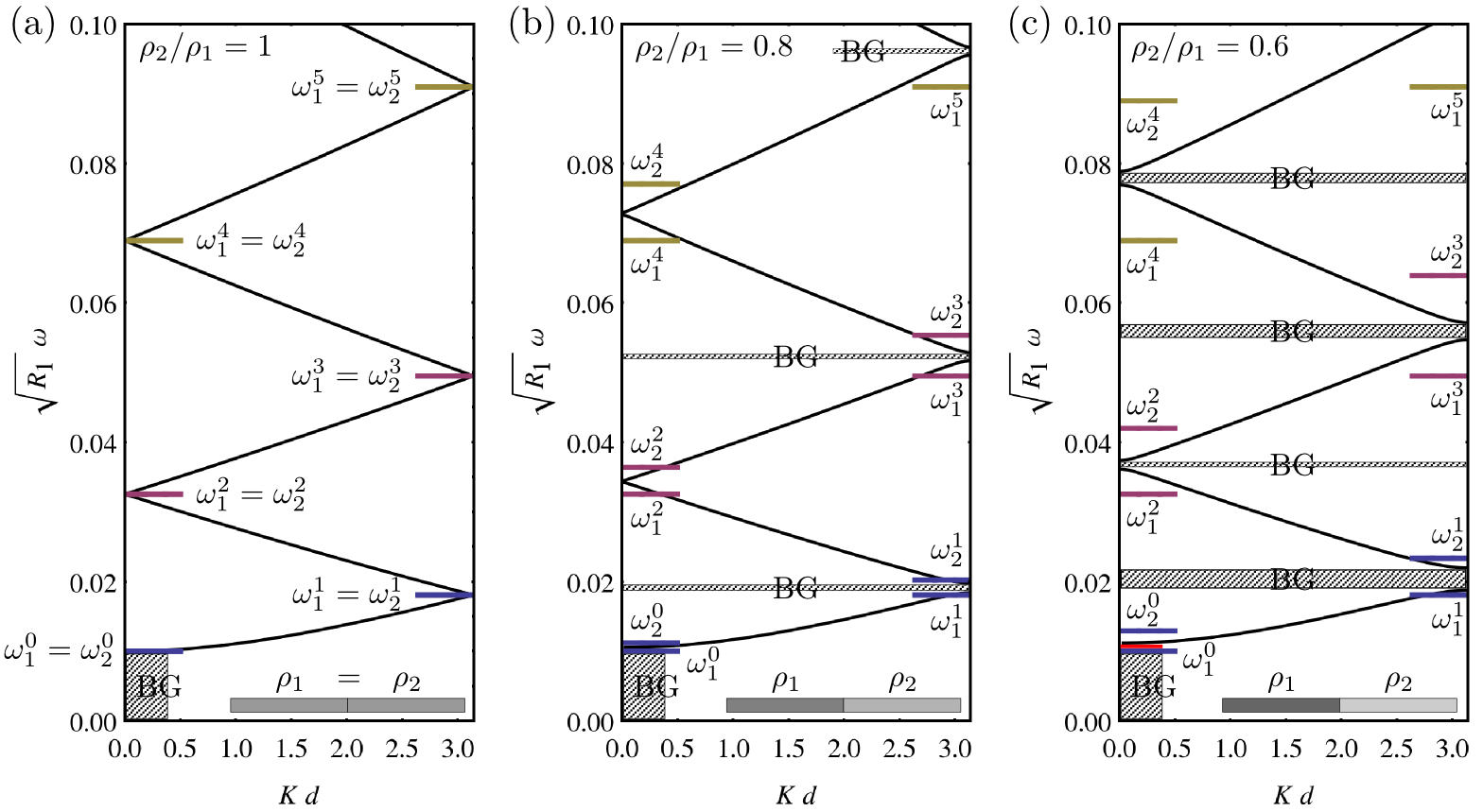}
\caption{\footnotesize Dispersion diagrams, representing a normalised circular  frequency $\sqrt{R_1}\omega$ vs.\ Bloch parameter 
$K d$, for a Rayleigh beam on an elastic foundation with piecewise constant mass density ($E_1 = E_2$, $A_1 = A_2$, $I_1 = I_2$, $\overline{P}_1 = \overline{P}_2 = 0.1$, $\overline{B}_1 = \overline{B}_2 = 0.0001$, $r/d = 0.015$, $l_1 = l_2 = d/2$): (a) Homogeneous beam $\rho_1 = \rho_2$ (for the homogeneous beam the dispersion diagram has been made artificially periodic by shifting the curve by multiples of $2\pi$); (b) Periodic beam with density contrast $\rho_2/\rho_1 = 0.8$; (c) Periodic beam with density contrast $\rho_2/\rho_1 = 0.6$. Horizontal segments indicate the predicted intersections $\omega_m^n$, see eq.\ (\ref{intersec}), for homogeneous beams with mass density either $\rho_1$ or $\rho_2$. By taking the midpoint between the horizontal segments, a good prediction of the band gap positions is obtained, especially for small contrast.}
\label{fig03}
\end{figure}

For each of these homogeneous beams, the dispersion curve is unique and is given by eq.\ (\ref{disp}).
Now, we shift the dispersion curve by multiples of $2\pi$, so that the dispersion diagram becomes artificially periodic, see Fig.~\ref{fig03}a. The intersections of the shifted curves are given by 
\begin{equation}
\label{intersec}
\omega_m^n = \sqrt{\frac{(\frac{n\pi}{d})^2r_m^2[(\frac{n\pi}{d})^2r_m^2 + \overline{P}_m] + \overline{B}_m}{R_m[(\frac{n\pi}{d})^2r_m^2 + 1]}}, \quad n = 0,1,2,3,...
\end{equation}
Note that for even values of $n$, we obtain the intersection points occurring at $K d = 0$, whereas for odd values of $n$, intersections occur at $K d = \pi$.

The formula (\ref{intersec}) can be used to estimate the position of the band gaps, as it is shown in Fig.~\ref{fig03} for the case of two constituents, having the same stiffness, $E_1 = E_2$, and cross-section, $A_1 = A_2$, $I_1 = I_2$, but different mass density (with contrast $\rho_2/\rho_1 = 0.8$  and $0.6$ in Fig.~\ref{fig03}b and \ref{fig03}c, respectively). In this example, the normalized prestress is $\overline{P}_1 = \overline{P}_2 = 0.1$ and the normalized Winkler stiffness $\overline{B}_1 = \overline{B}_2 = 0.0001$. The unit cell is chosen such that $r/d = 0.015$ and $l_1 = l_2 = d/2$.

It is clear from the figure that, as the contrast in mass density increases, band gaps open up and, at the same time, they move upward. The predicted intersections for homogeneous beams having either mass density $\rho_1$ and $\rho_2$ also open up. These intersections are indicated in the figure by horizontal segments. Note that the segment corresponding to phase 1 remains fixed, while the segment corresponding to phase 2 move upward.
By taking the midpoint between two corresponding horizontal segments, it is possible to obtain a good prediction of the band gap positions, especially for small contrast.

\newpage

\section{Green's function for a Rayleigh beam. Localised and propagating wave forms}
\label{sec04}

In this section, we first construct the time-harmonic Green's function associated to a concentrated unit force applied to a infinite, homogeneous, prestressed Rayleigh beam on an elastic foundation. Also, we derive the solution associated to a concentrated couple. 

On the basis of the fundamental solutions, we analyse localised wave forms (produced by a rigid body attached to a Rayleigh beam) as well as propagating modes.

\subsection{Concentrated force}

The time-harmonic Green's function $\mG_F(x,t) = G_F(x) \exp(-i\omega t)$ corresponding to a transverse concentrated unit force placed at the origin $x = 0$ of a homogeneous Rayleigh beam (assumed positive when oriented as the displacement $v$) solves the equation
\begin{equation}
\label{delta}
G_F''''(x) - 2b\, G_F''(x) + c\, G_F(x) = \frac{\delta(x)}{EI},
\end{equation}
where $\delta(x)$ is the Dirac delta function and 
\begin{equation}
b = \frac{\overline{P}}{2r^2} - \frac{R}{2r^2}\omega^2, \quad
c = \frac{\overline{B}}{r^4} - \frac{R}{r^4}\omega^2.
\end{equation}
Assuming that $c \neq 0$ and $b^2 - c \neq 0$, a solution of the corresponding homogeneous equation can be sought in the form $C \exp(ikx)$, which leads to the characteristic equation
\begin{equation}
\label{eqcar}
k^4 + 2b\, k^2 + c = 0.
\end{equation}

\paragraph{Stop band Green's function.}
Assuming a \lq forbidden' frequency $\omega$ within a band-gap, the four roots of the characteristic equation (\ref{eqcar}) possess non-zero imaginary parts, corresponding to a localised (or evanescent) mode.
We distinguish between two cases:
\begin{itemize}

\item Case 1:  Low frequency and high tensile prestress ($c \geq 0$, $b > 0$, $b^2 - c > 0$).

The four roots are purely imaginary and admit the following representation
\begin{equation}
k = \pm i k_{I_{1,2}}, \quad k_{I_{1,2}} = \sqrt{b \pm \sqrt{b^2 - c}}.
\end{equation}
In this case, the Green's function takes the form
\begin{equation}
G_F(x) = -\frac{1}{2EI(k_{I_1}^2 - k_{I_2}^2)} \left\{ \frac{1}{k_{I_1}}e^{-k_{I_1}|x|} - 
\frac{1}{k_{I_2}}e^{-k_{I_2}|x|} \right\},
\end{equation}
which corresponds to a monotonic decaying mode, without oscillations.

\item Case 2:  Low frequency and low prestress (tensile or compressive) ($c > 0$, $b^2 - c < 0$).

The four roots are complex and admit the following representation
\begin{equation}
k = \pm (k_R \pm i k_{I}), \quad k_{R} = \sqrt{\frac{\sqrt{c} - b}{2}}, \quad 
k_{I} = \sqrt{\frac{\sqrt{c} + b}{2}}.
\end{equation}
In this case, the Green's function takes the form
\begin{equation}
G_F(x) = \frac{e^{-k_I |x|}}{4EI(k_R^2 + k_I^2)}\left\{ \frac{1}{k_I}\cos(k_R|x|) + 
\frac{1}{k_R}\sin(k_R|x|) \right\},
\end{equation}
which corresponds to a oscillating decaying mode.

\end{itemize}

\paragraph{Pass band Green's function.}
Assuming a frequency $\omega$ within a pass band, the characteristic equation (\ref{eqcar}) always admits real roots corresponding to a propagating wave. We distinguish between two cases:

\begin{itemize}

\item Case 3: Low frequency and high compressive prestress ($c \geq 0$, $b^2 - c > 0$, $b < 0$).

The four roots are real and admit the following representation
\begin{equation}
k = \pm k_{R_{1,2}}, \quad k_{R_{1,2}} = \sqrt{-b \pm \sqrt{b^2 - c}}.
\end{equation}
In this case the Green's function takes the form
\begin{equation}
G_F(x) = \frac{i}{2EI (k_{R_1}^2 - k_{R_2}^2)} \left\{ \frac{1}{k_{R_1}} e^{ik_{R_1}|x|} - \frac{1}{k_{R_2}} e^{ik_{R_2}|x|} \right\},
\end{equation}
which corresponds to two waves travelling away from the source ({\em outgoing waves}: the {\em Sommerfeld radiation condition} states that the wave must travel from the point of application of the concentrated force to infinity).

\item Case 4: High frequency and any prestress ($c < 0$, $b^2 - c > 0$).

The four roots are real and imaginary, and admit the following representation
\begin{alignat}{2}
k &= \pm k_R, & \quad k_R &= \sqrt{-b + \sqrt{b^2 - c}}, \label{eq37} \\
k &= \pm i k_I, & \quad k_I &= \sqrt{b + \sqrt{b^2 - c}}. \label{eq38} 
\end{alignat}
In this case the Green's function takes the form
\begin{equation}
G_F(x) = \frac{1}{2EI (k_R^2 + k_I^2)} \left\{ \frac{i}{k_R} e^{ik_R |x|} - \frac{1}{k_I} e^{-k_I |x|} \right\},
\end{equation}
which corresponds to a evanescent wave superimposed to an {\em outgoing} travelling wave. 

\end{itemize}

\begin{figure}[!htcb]
\centering
\includegraphics[width=6.5cm]{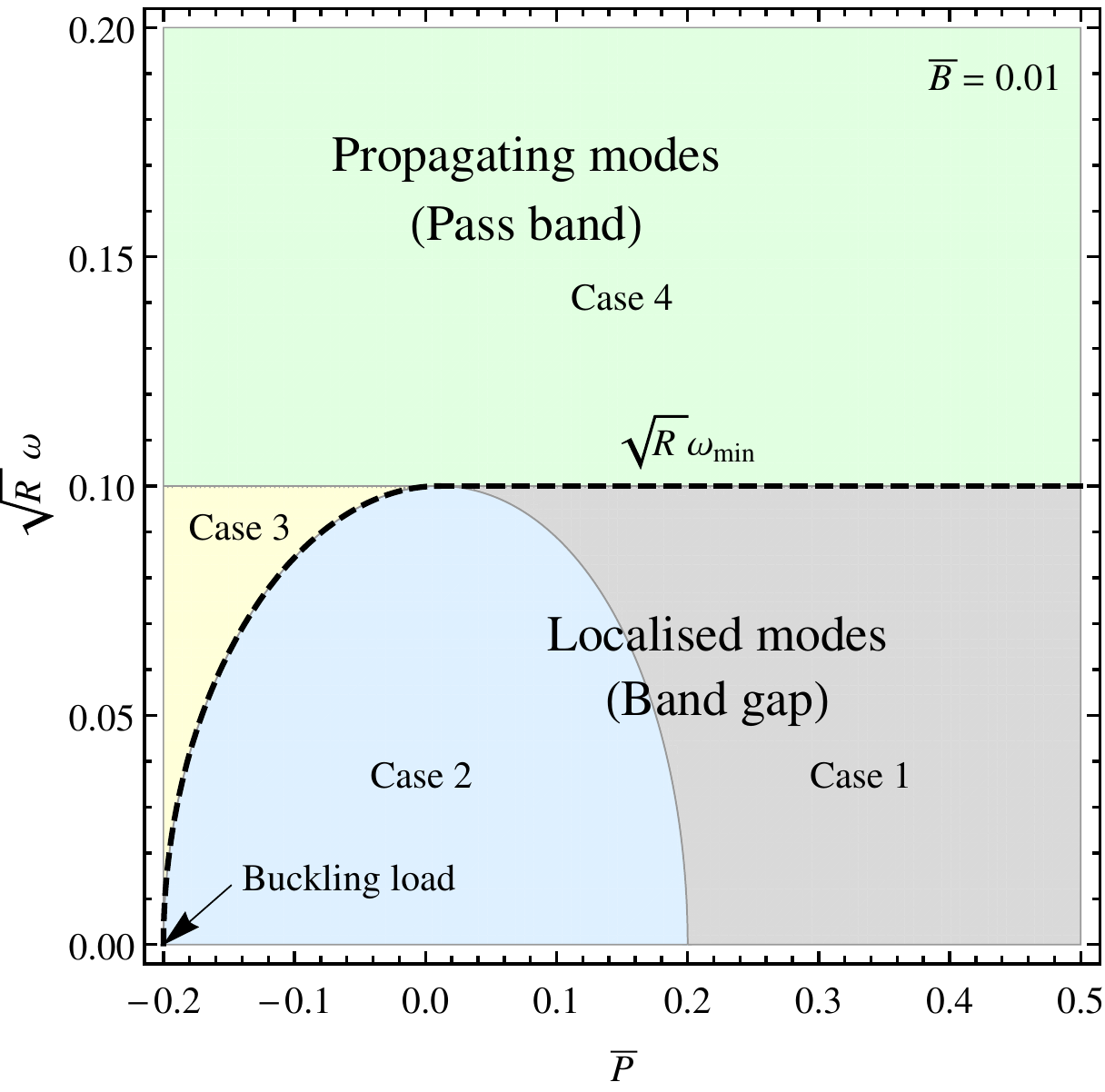}
\caption{\footnotesize Map of four dynamic regimes for an infinite, homogeneous, prestressed Rayleigh beam on an elastic foundation in the plane $\overline{P}$ -- $\sqrt{R}\, \omega$ ($\overline{B} = 0.01$). Localised modes correspond to a frequency within a band gap (below $\sqrt{R}\, \omega_\text{min}$, reported with a dashed line): monotonic decay without oscillations (Case 1); decay with oscillations (Case 2). Propagating modes correspond to a frequency within a pass band: two travelling waves (Case 3); evanescent wave superimposed to a travelling wave (Case 4).}
\label{cases}
\end{figure}

The four regimes (Cases 1--4) are shown in Fig.~\ref{cases} in the plane $\overline{P}$ -- $\sqrt{R}\, \omega$ for foundation stiffness $\overline{B} = 0.01$. Localised wave forms are observed for frequencies within a band gap (below $\sqrt{R}\, \omega_\text{min}$, reported with a dashed line): Case 1 corresponds to a monotonic decay (without oscillations); Case 2 corresponds to an oscillatory decay. On the other hand, propagating modes are observed for frequencies within a pass band: Case 3 corresponds to two travelling waves; Case 4 corresponds to an evanescent wave superimposed to a travelling wave.

\paragraph{Special resonant cases giving rise to linear or quasi-linear growth at infinity.}

Special cases are obtained for the choice of dynamic parameters for which $c = 0$ or $b^2 - c = 0$. In these cases, the characteristic equation (\ref{eqcar}) admits double roots, resulting in the linear or quasi-linear growth at infinity, as shown in Fig.~\ref{fig00} in the introduction.

\begin{itemize}

\item Special case $c = 0$, $b > 0$

\begin{equation}
G_F(x) = -\frac{1}{2EI k_I^3} \left\{ e^{-k_I |x|} + k_I |x| \right\}, \label{GF40}
\end{equation}
where $k_I = \sqrt{2b}$.

\item Special case $c = 0$, $b < 0$

\begin{equation}
G_F(x) = \frac{1}{2EI k_I^3} \left\{ i e^{i k_R |x|} + k_R |x| \right\}, \label{GF41}
\end{equation}
where $k_R = \sqrt{-2b}$.

\item Special case $b^2 - c = 0$, $b < 0$

\begin{equation}
G_F(x) = -\frac{1}{4EI k_R^3} e^{i k_R |x|} \left\{ i  + k_R |x| \right\}, \label{GF42}
\end{equation}
where $k_R = \sqrt{-b}$.

\end{itemize}

\subsection{Concentrated couple}

The time-harmonic function $\mG_C(x,t) = G_C(x) \exp(-i\omega t)$ corresponding to a concentrated unit couple (assumed positive when oriented as the rotation $v'$) placed at the origin $x = 0$ of a homogeneous Rayleigh beam is the solution of the following equation
\begin{equation}
\label{deltap}
G_C''''(x) - 2b\, G_C''(x) + c\, G_C(x) = -\frac{\delta'(x)}{EI}.
\end{equation}
Note that the function $G_C(x)$ has the property $G_C(x) = -G_F'(x)$. For the four cases introduced above, the function $G_C$ takes the form
\begin{itemize}

\item Case 1:  Low frequency and high tensile prestress ($c \geq 0$, $b > 0$, $b^2 - c > 0$).

\begin{equation}
G_C(x) = -\frac{\sign(x)}{2EI(k_{I_1}^2 - k_{I_2}^2)} \left\{ e^{-k_{I_1}|x|} - e^{-k_{I_2}|x|} \right\},
\end{equation}
which corresponds to a monotonic decaying mode, without oscillations.

\item Case 2:  Low frequency and low prestress (tensile or compressive) ($c > 0$, $b^2 - c < 0$).

\begin{equation}
G_C(x) = \frac{e^{-k_I |x|}}{4EI k_R k_I} \sin(k_R x),
\end{equation}
which corresponds to a oscillating decaying mode.

\item Case 3: Low frequency and high compressive prestress ($c \geq 0$, $b^2 - c > 0$, $b < 0$).

\begin{equation}
G_C(x) = \frac{\sign(x)}{2EI (k_{R_1}^2 - k_{R_2}^2)} \left\{ e^{ik_{R_1}|x|} - e^{ik_{R_2}|x|} \right\}.
\end{equation}
which corresponds to two outgoing travelling waves.

\item Case 4: High frequency and any prestress ($c < 0$, $b^2 - c > 0$).

\begin{equation}
G_C(x) = \sign(x) \frac{1}{2EI (k_R^2 + k_I^2)} \left\{ e^{ik_R |x|} - e^{-k_I |x|} \right\}.
\end{equation}
which corresponds to a evanescent wave superimposed to an {\em outgoing} travelling wave. 

\end{itemize}

\subsection{Localised wave forms}

\begin{figure}[!htcb]
\centering
\includegraphics[width=13cm]{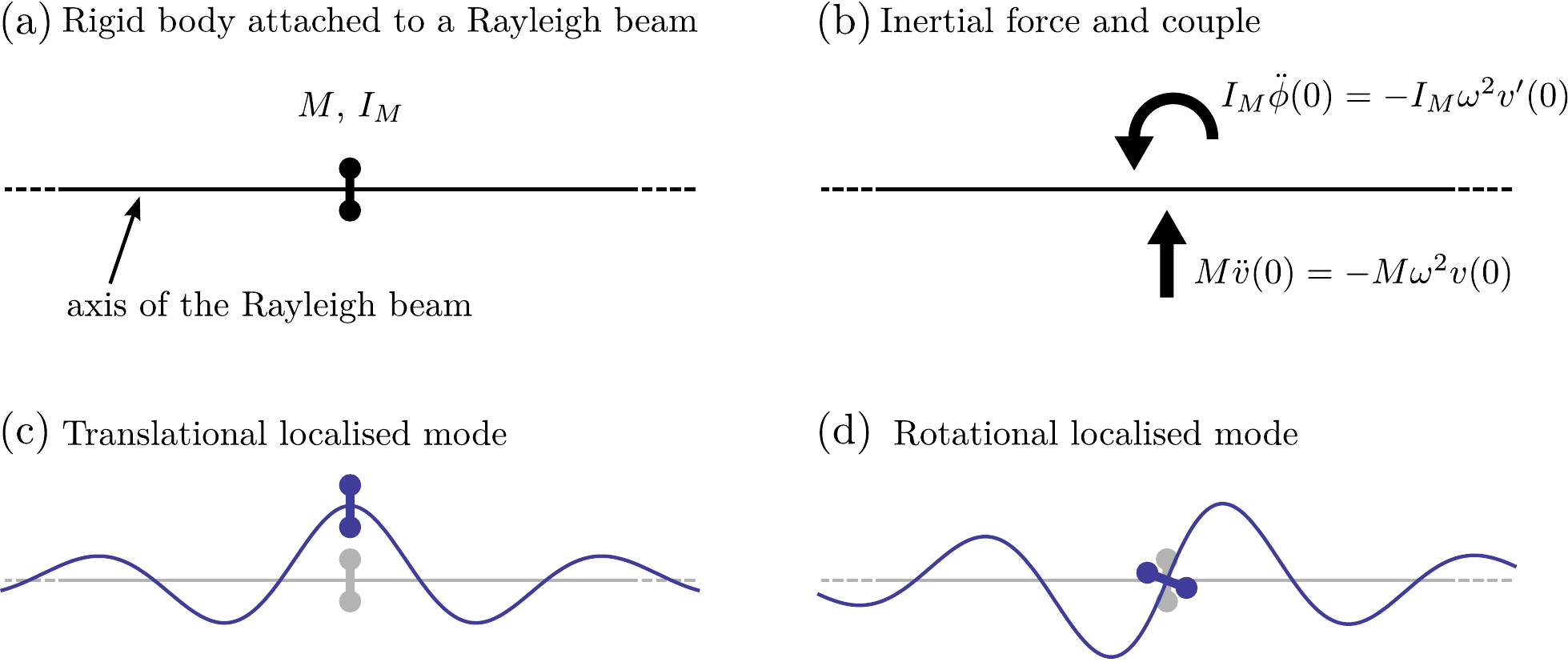}
\caption{\footnotesize (a) A rigid body having mass $M$ and moment of inertia $I_M$ attached to a Rayleigh beam at $x = 0$. (b) Inertial force $M \ddot{v}(0)$ and inertial couple $I_M \ddot{\phi}(0)$ transmitted by the body to the beam. (c) Translational localised mode corresponding to $v(0) \neq 0$ and $v'(0) = 0$. (d) Rotational localised mode corresponding to $v(0) = 0$ and $v'(0) \neq 0$.}
\label{mass}
\end{figure}

We now consider a rigid body, having the mass $M$ and the moment of inertia $I_M$, attached to the beam at the origin and vibrating at a frequency $\omega$ within a band gap, see Fig.~\ref{mass}a. The effect of the rigid body is replaced by the inertial force $M \ddot{v}(0) = -M \omega^2 v(0)$ and the inertial couple $I_M \ddot{\phi}(0) = -I_M \omega^2 v'(0)$, see Fig.~\ref{mass}b. 

The transverse displacement $v(x)$ is given in terms of Green's functions as follows
\begin{equation}
\label{emme1}
v(x) = M \omega^2 v(0) G_F(x) + I_M \omega^2 v'(0) G_C(x).
\end{equation}
By taking the derivative of the previous equation, we get
\begin{equation}
\label{emme2}
v'(x) = M \omega^2 v(0) G_F'(x) + I_M \omega^2 v'(0) G_C'(x).
\end{equation}
Eqs. (\ref{emme1}) and (\ref{emme2}) can then be evaluated at $x = 0$, which leads to the system
\begin{equation}
\left\{
\begin{aligned}
v(0) &= M \omega^2 v(0) G_F(0) + I_M \omega^2 v'(0) G_C(0), \\[3mm]
v'(0) &= M \omega^2 v(0) G_F'(0) + I_M \omega^2 v'(0) G_C'(0).
\end{aligned}
\right.
\end{equation}
For the Rayleigh beam, we have $G_F'(0) = G_C(0) = 0$, so that the system decouples and we obtain the 
equations
\begin{equation}
\label{system}
M = \frac{1}{\omega^2 G_F(0)}, \quad 
I_M = \frac{1}{\omega^2 G_C'(0)}.
\end{equation}
We note that two independent localised modes are identified, which will be referred to as a {\em translational mode} and a {\em rotational mode}, see Fig.~\ref{mass}c and \ref{mass}d, respectively. The {\em translational localised mode} is obtained by setting $v(0) \neq 0$ and $v'(0) = 0$; in this case the rotation at the origin is zero, so that the moment of inertia $I_M$ is not involved and the motion is sustained only by the mass $M$ given by the first equation in (\ref{system}). The {\em rotational localised mode} is obtained by setting $v(0) = 0$ and $v'(0) \neq 0$; in this case the transverse displacement at the origin is zero, so that the mass $M$ is not involved and the motion is sustained only by the moment of inertia $I_M$ given by the second equation in (\ref{system}).

\begin{figure}[!htcb]
\centering
\includegraphics[width=15cm]{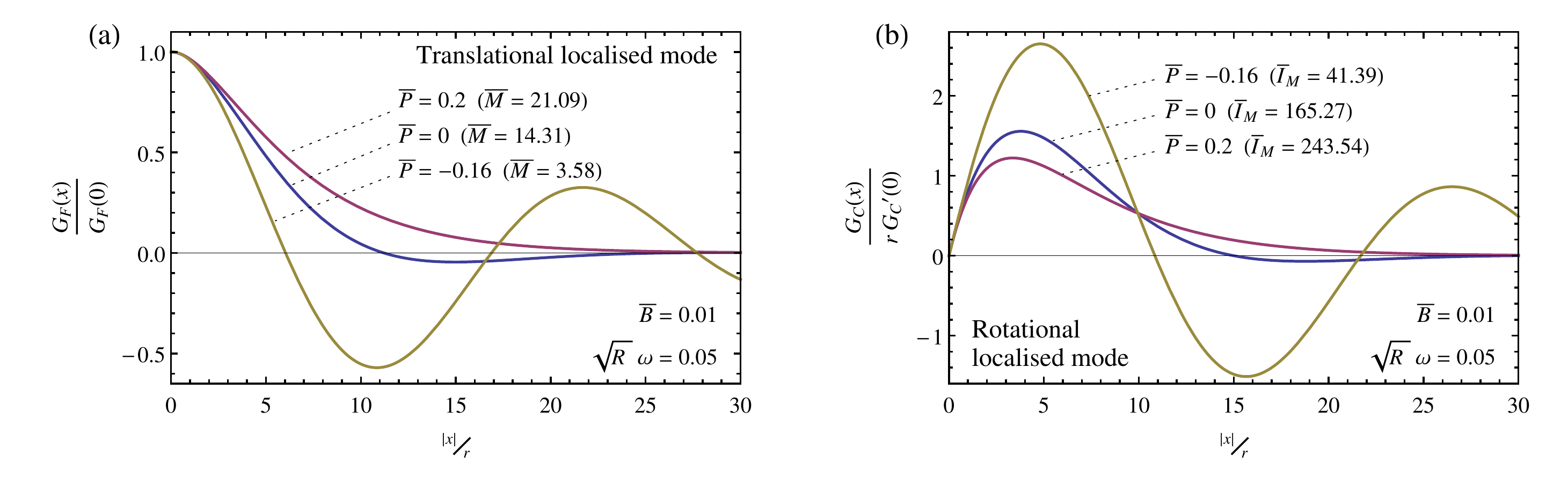}
\caption{\footnotesize Amplitude of localised vibration modes along the $x$-axis of the beam for normalised frequency $\sqrt{R}\, \omega = 0.05$, foundation stiffness $\overline{B} = 0.01$, and different values of prestress $\overline{P} = \{ -0.16, 0, 0.2 \}$. (a) Translational localised mode, the amplitude is normalised by the value at the origin. The mass $\overline{M}$ is located at $x = 0$. Note that the rotation at the origin is zero for this mode and thus the moment of inertia $\overline{I}_M$ is not involved. (b) Rotational localised mode, the amplitude is normalised by $r$ and the value of rotation at the origin. The rigid body having moment of inertia $\overline{I}_M$ is located at $x = 0$. Note that the transverse displacement at the origin is zero for this mode and thus the mass $\overline{M}$ is not involved.}
\label{figloc2}
\end{figure}

Examples of translational and rotational localised modes are shown in Fig.~\ref{figloc2}a and \ref{figloc2}b, respectively, for normalised frequency $\sqrt{R}\, \omega = 0.05$, foundation stiffness $\overline{B} = 0.01$, and different values of prestress $\overline{P} = \{ -0.16, 0, 0.2 \}$. The value of normalised mass $\overline{M} = M / (2\rho A r)$ (for translational mode) and normalised moment of inertia $\overline{I}_M = I_M / (2\rho I r)$ (for rotational mode) needed to sustain the motion for each value of prestress is indicated in the figure.

\begin{figure}[!htcb]
\centering
\includegraphics[width=15cm]{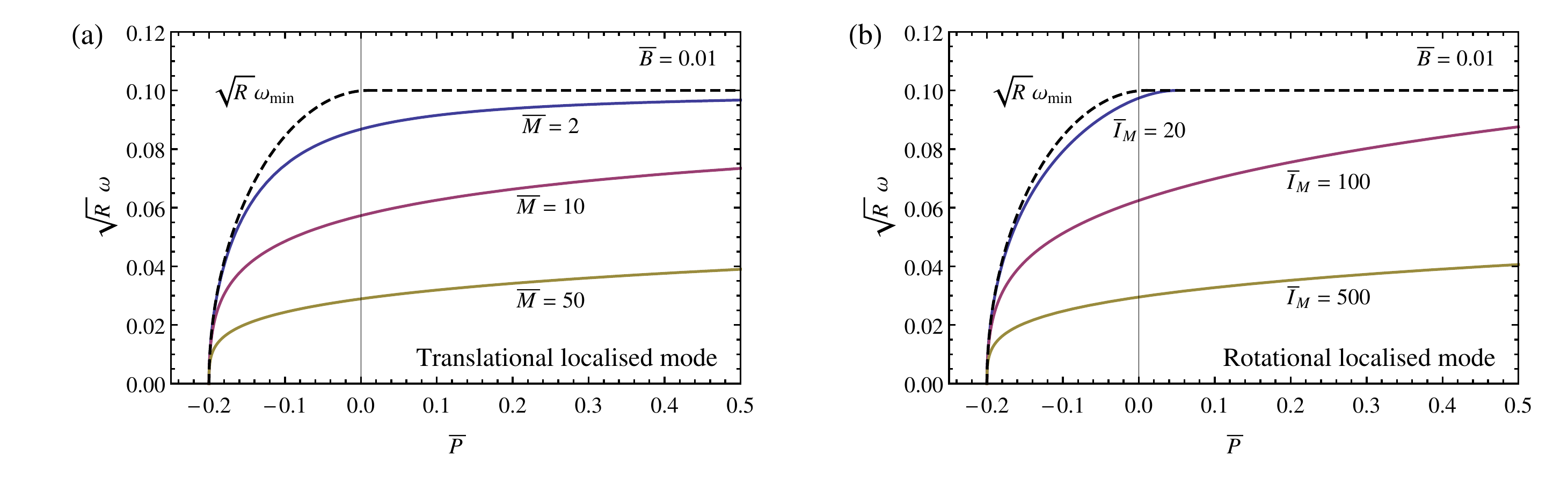}
\caption{\footnotesize Normalised circular frequency $\sqrt{R}\, \omega$ associated to a body, having dimensionless mass $\overline{M}$ and dimensionless moment of inertia $\overline{I}_M$, attached to an infinite, homogeneous, prestressed Rayleigh beam on an elastic foundation as a function of prestress $\overline{P}$ ($\overline{B} = 0.01$). The upper boundary of the band gap $\sqrt{R}\, \omega_\text{min}$ is reported as a dashed line. (a) Translational localised mode corresponding to $v(0) \neq 0$, $v'(0) = 0$, sustained by the mass $\overline{M}$. (b) Rotational localised mode corresponding to $v(0) = 0$, $v'(0) \neq 0$, sustained by the moment of inertia $\overline{I}_M$.}
\label{figloc}
\end{figure}

\noindent
Explicit formulae can be obtained for the mass $M$ and the moment of inertia $I_M$ associated to the translational and rotational localised mode, namely
\begin{equation}
\begin{aligned}
M &= \frac{4EI k_I(k_R^2 + k_I^2)}{\omega^2} = \frac{2EI}{r^3 \omega^2} 
\sqrt{\overline{B} - R \omega^2} \sqrt{\overline{P} - R\omega^2 + 
2\sqrt{\overline{B} - R\omega^2}}, \\[3mm]
I_M &= \frac{4EI k_I}{\omega^2} = \frac{2EI}{r \omega^2} \sqrt{\overline{P} - R\omega^2 + 
2\sqrt{\overline{B} - R\omega^2}}.
\end{aligned}
\end{equation}
The dependency of the dimensionless frequency $\sqrt{R}\, \omega$ upon the prestress $\overline{P}$ for $\overline{B} = 0.01$ and three values of $\overline{M}$ (translational mode) and three values of $\overline{I}_M$ (rotational mode) is reported in Fig.~\ref{figloc}a and \ref{figloc}b, respectively.

\subsection{Propagating wave forms}

\begin{figure}[!htcb]
\centering
\includegraphics[width=15cm]{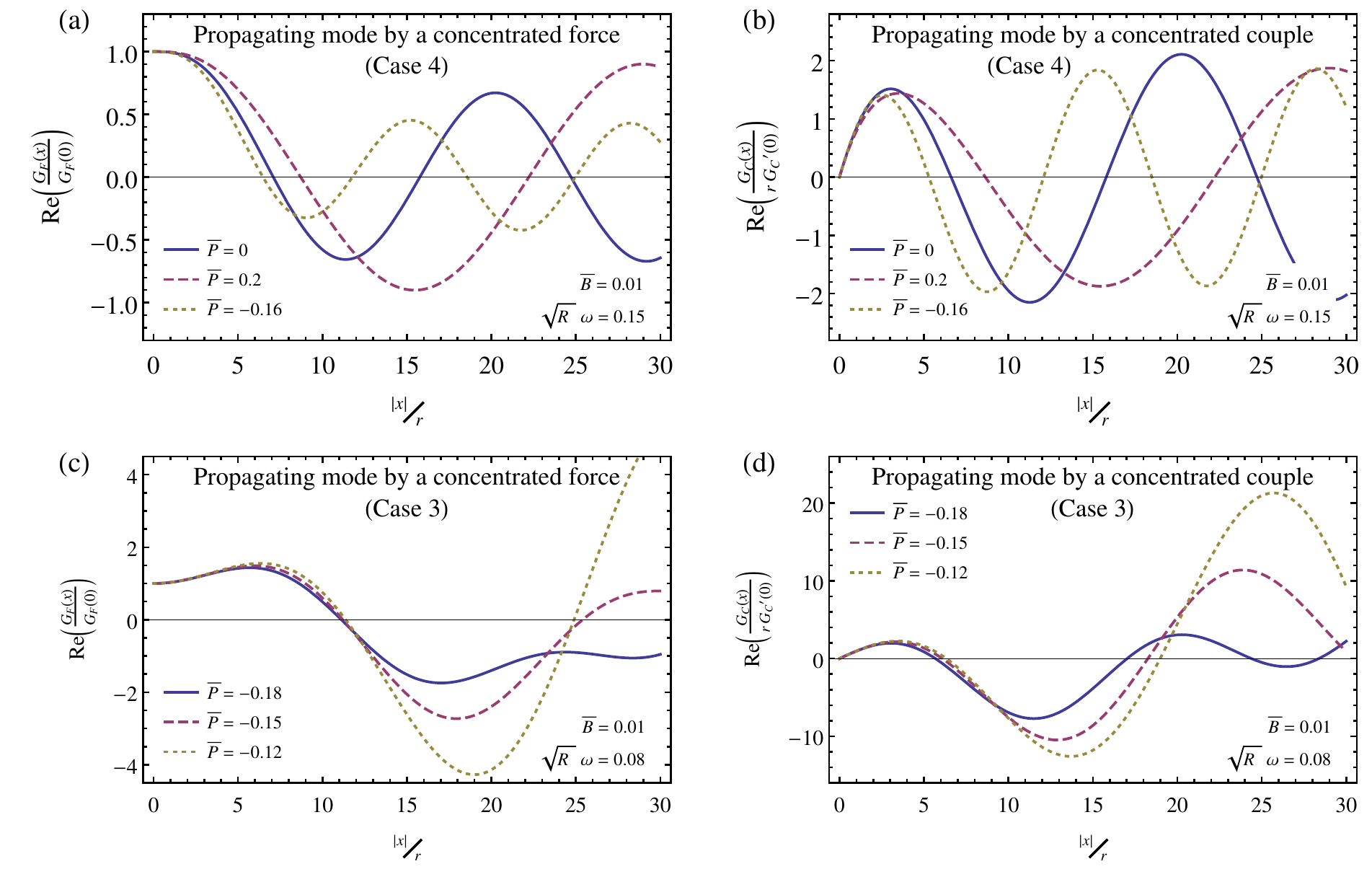}
\caption{\footnotesize Amplitudes of propagating modes along the $x$-axis of the beam for Winkler stiffness $\overline{B} = 0.01$. Propagating  mode induced by a concentrated force (a), and a concentrated couple (b), placed at the origin for normalised frequency $\sqrt{R}\, \omega = 0.15$ (high-frequency, Case 4) and different values of prestress $\overline{P} = \{ -0.16, 0, 0.2 \}$. Propagating  mode induced by a concentrated force (c), and a concentrated couple (d), placed at the origin for normalised frequency $\sqrt{R}\, \omega = 0.08$ (low-frequency and high compressive prestress, Case 3) and different values of prestress $\overline{P} = \{ -0.18, -0.15, -0.12 \}$.}
\label{propagm}
\end{figure}

Propagating wave forms for the case of high frequency and different values of prestress (Case 4) and the case of low frequency and high compressive prestress (Case 3) are shown in Fig.~\ref{propagm}.

These solutions are bounded at infinity, which is different from the special resonant cases of Green's function for the Rayleigh beam on the Winkler foundation, as shown in Fig.~\ref{fig00} in the introduction, where the neighbourhood of the origin was an effective hinge dominated by an exponentially localised wave form, whereas the linear growth was present at infinity.
In parts (a) and (b), we show the Green's function (point force solution) and its derivative (concentrated couple) for the case when the solution includes a propagating wave as well as an evanescent term. This is reflected in the higher amplitude at the origin, as observed in the Fig.~\ref{propagm}a.
On the contrary, the parts (c) and (d) include the Green's function and its derivative for the choice of dynamic parameters giving rise to two propagating waves, without an evanescent term present. The wave pattern is typical for interference of two periodic solutions, as expected in this case.   

The pass band Green's function will be used in the next section in order to construct the quasi-periodic solutions corresponding to a multi-mass system in a Rayleigh beam.

\clearpage

\section{Quasi-periodic Greens function. Dispersion of Bloch waves for a multi-mass system in a Rayleigh beam}
\label{sec05}

The results of Section \ref{sec04} are extended here to the cases of periodic multi-mass systems. An infinite Rayleigh beam is taken together with a periodic array of identical bodies possessing mass $M$ and the moment of inertia $I_M$. Instead of a single source or a localised couple, discussed in the earlier section \ref{sec04}, we now have a periodic array of point forces and point moments accordingly.
For the purpose of analysis of Bloch waves in the periodic structured Rayleigh beam, we need the quasi-periodic Green's function.

Assuming that $G_F(x; \omega)$ is a pass-band Green's function for a single point force applied to the origin at the Rayleigh beam, the quasi-periodic Green's function $g_F(x; \omega, K)$ representing a periodically distributed sources generating a propagating wave, has the form 
\begin{equation}
g_F(x; \omega, K) = \sum_{n=-\infty}^{\infty} G_F(x - nd; \omega) e^{i K nd},
\end{equation}
where $d$ stands for the period of the structure, and in the computations below we assume $d = 1$.

It is also noted that the function $g_F$ satisfies the differential equation:
\begin{equation}
g_F'''' - 2b\, g_F'' + c\, g_F = \sum_{n=-\infty}^{\infty} \frac{\delta(x - nd)}{EI} e^{i K nd},
\end{equation}
with delta function terms being periodically distributed along the infinite beam, and the phase-shift factors being applied accordingly to the forcing terms.

Along the same lines, a function $g_C(x; \omega, K)$ is introduced as the solution of the equation
\begin{equation}
g_C'''' - 2b\, g_C'' + c\, g_C = -\sum_{n=-\infty}^{\infty} \frac{\delta'(x - nd)}{EI} e^{i K nd},
\end{equation}
which represents an infinite array of periodically distributed concentrated moments subjected to the 
Floquet phase shift accordingly. We note that $g_C$ can be written in terms of the single couple Green's function $G_C(x;\omega)$ as follows
\begin{equation}
g_C(x; \omega, K) = \sum_{n=-\infty}^{\infty} G_C(x - nd; \omega) e^{i K nd}.
\end{equation}
It is also straightforward to see that
\begin{equation}
g_C(x; \omega, K) = -g_F'(x; \omega, K).
\end{equation}
The flexural displacement $v(x)$ in the Rayleigh beam, subjected a action of periodically distributed forces and moments, satisfies the following equations
\begin{equation}
v(x) = m \omega^2 v(0) g_F(x; \omega, K) + I_M \omega^2 v'(0) g_C(x; \omega, K)
\end{equation}
and
\begin{equation}
v'(x) = m \omega^2 v(0) g_F'(x; \omega, K) + I_M \omega^2 v'(0) g_C'(x; \omega, K).
\end{equation}
Evaluation of $v$ and $v'$ at the origin leads to the homogeneous system of linear algebraic equations
\begin{equation}
\begin{bmatrix}
1 - M \omega^2 g_F(0;\omega, K) & -I_M \omega^2 g_C(0;\omega, K) \\
- M \omega^2 g_F'(0;\omega, K)  & 1 - I_M \omega^2 g_C'(0;\omega, K)
\end{bmatrix}
\begin{bmatrix}
v(0) \\
v'(0)
\end{bmatrix} = 
\begin{bmatrix}
0 \\
0
\end{bmatrix}.
\end{equation}
This system has non-trivial solutions if and only if the determinant of its matrix is zero, i.e.
\begin{equation}
\label{disp2}
1 - \omega^2 (Mg_F(0;\omega, K) + I_M g_C'(0;\omega, K)) + MI_M \omega^4 (g_F(0;\omega, K)g_C'(0;\omega, K) -g_F'(0;\omega, K)g_C(0;\omega, K)) = 0.
\end{equation}
The above equation (\ref{disp2}) defines $\omega$ as an implicit function of $K$ and hence it is the dispersion equation for Bloch-Floquet waves in the structured Rayleigh beam.

\begin{figure}[!htcb]
\centering
\includegraphics[width=15cm]{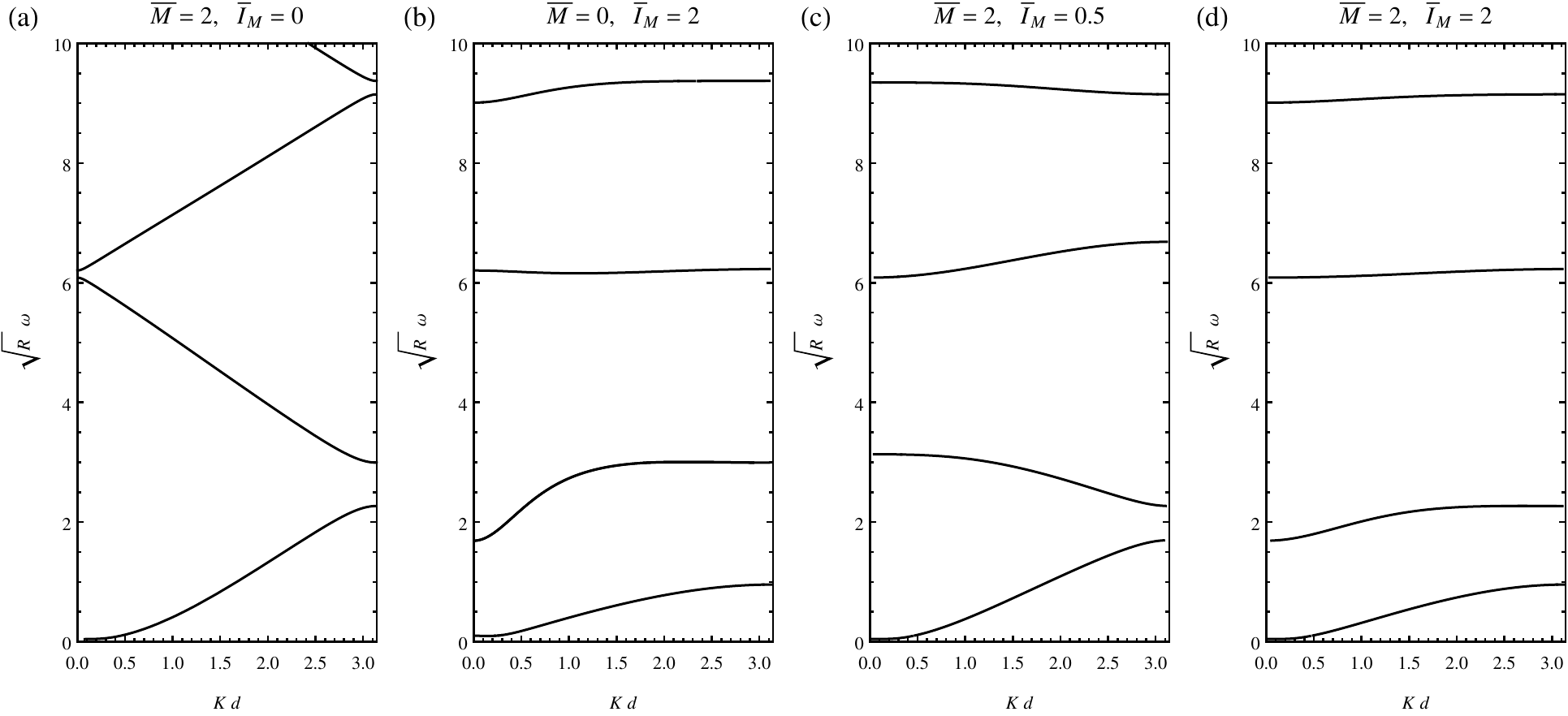}
\caption{\footnotesize Dispersion diagrams for a multi-mass system in a Rayleigh beam (for foundation stiffness $\overline{B} = 0.01$, null prestress $\overline{P} = 0$ and unit cell size $d = r$). $\overline{M} = M / (2\rho A r)$ and $\overline{I}_M = I_M / (2\rho I r)$ are the dimensionless mass and moment of inertia of the bodies periodically distributed along the beam.}
\label{figQPj}
\end{figure}

We note that the nodal values of the quasi-periodic Green's functions $g_F(0;\omega, K)$ and $g_C(0;\omega, K)$, used in \eq{disp2}, are evaluated in the explicit form, 
\begin{align}
g_F(0;\omega, K) &= \frac{1}{2EI k_R k_I (k_R^2 + k_I^2)}
\left(
k_R \frac{\sinh(k_I d)}{\cos(K d) - \cosh(k_I d)} - 
k_I \frac{\sin(k_R d)}{\cos(K d) - \cos(k_R d)}
\right), \label{GF} \\[3mm]
g_C(0;\omega, K) &= -g_F'(0;\omega, K) = \frac{i \sin(K d)}{2EI (k_R^2 + k_I^2)} \cdot
\left(
\frac{1}{\cos(K d) - \cos(k_R d)} -
\frac{1}{\cos(K d) - \cosh(k_I d)}
\right), \label{GF1} \\[3mm]
g_C'(0;\omega, K) &= -g_F''(0;\omega, K) = -\frac{1}{2EI (k_R^2 + k_I^2)}
\left(
k_R \frac{\sin(k_R d)}{\cos(K d) - \cos(k_R d)} + 
k_I \frac{\sinh(k_I d)}{\cos(K d) - \cosh(k_I d)}
\right), \label{GF2}
\end{align}
where $k_R$ and $k_I$ are defined in (\ref{eq37}) and (\ref{eq38}). With reference to \eq{GF}--\eq{GF2}, we note that all the coefficients of the equation \eq{disp2} are real, and the dispersion diagrams (see Fig.~\ref{figQPj}) can be constructed by obtaining real roots $K$ of \eq{disp2} for given values of the radian frequency $\omega$. It also worthwhile saying that the left-hand side of \eq{disp2} is a rational function of $Z = \cos(Kd)$.

Furthermore, the dispersion equation \eq{disp2} is solved 
to demonstrate the dispersion properties of Bloch waves in a structured Rayleigh beam.

Fig.~\ref{figQPj} incorporates four examples, all showing band gaps within the low frequency range.
In particular, the diagram (a) of this figure corresponds to the case when point masses, without rotational inertia, are distributed periodically along the length of the beam, whereas the diagram (b) describes the case when the mass is formally taken as zero and only rotational inertia is taken into account. While for low frequencies the diagram (a) is similar to the one observed in the Euler--Bernoulli structured beam, the second case (b) appears to be important for the effects of the rotational inertia, and in particular, it shows waves of small group velocity and wide band gaps in the low frequency range.

In the next two diagrams (c) and (d) of Fig.~\ref{figQPj}, we show the combined effects of translational and rotational inertia produced by bodies placed periodically along the length of the Rayleigh beam.
The comparison of these computations shows interesting localisation effects associated with the rotational inertia. With the increase of the rotational inertia of periodically distributed solids, the dispersion curves become flatter, which implies reduction in the absolute value of the group velocity. This effect is especially pronounced in the high frequency range. An observation of special importance is in the change of sign of the group velocity for the second band, as follows from the comparison of diagrams (c) and (d) for $\overline{I}_M=0.5$ and  $\overline{I}_M=2$, and also the change of the sign of curvature of the dispersion curve in the second band. 

\newpage

\section{Conclusion}
\label{sec06}

The paper has addressed an important class of problems that arise for flexural Bloch waves in elastic systems with 
rotational inertia and prestress, in the presence of a Winkler foundation. This has been done for so-called Rayleigh beams and structured Rayleigh beams. We have also highlighted a direct analogy with the models of couple-stress materials; although physical origins of the two problems are different, the mathematical underlay is identical and hence the mathematical study of the dynamic response of structured Rayleigh beams also brings a valuable insight into the study of shear waves in Cosserat type composites.   

The dispersion properties of waves in the Rayleigh beam appear to be very different from those in Euler beams in the high frequency regimes. This has been demonstrated analytically. The combined action of rotational inertia and prestress has been highlighted. The low frequency bandgaps have been created via introduction of the term that corresponds to an elastic foundation of the Winkler type. 

Defect modes and localisation have been studied through analysis of stop band and pass band dynamic Green's functions. Quasi-periodic Green's functions for translational and rotational inertia have been constructed, and the corresponding dispersion equations are written in closed form, which allows for effective and accurate solution and analysis.

The analysis presented in this work is generic and is straightforward to extend to flexural systems such as plates and shells, in the cases where rotational inertia effects appear to be important. This is likely to be applicable to the frame-type flexural systems as well as high contrast systems, modelling dynamic shields of elastic waves (also known as ''invisibility cloaks'') and earthquake protection systems.

\vspace{6mm}
{\bf Acknowledgements}. AP would like to acknowledge financial support from the
European Union's Seventh Framework Programme FP7/2007-2013/ under REA grant
agreement number PCIG13-GA-2013-618375-MeMic.
AM gratefully acknowledges financial support from the
European Union's Seventh Framework Programme FP7/2007-2013/ under REA grant
agreement number PITN-GA-2013-606878-CERMAT2.

\bibliographystyle{jabbrv_unsrt}
\bibliography
{%
../../BIBTEX/roaz}

%
%

\end{document}